\documentclass[9pt,twocolumn,twoside]{opticajnl}

\journal{ol} 

\setboolean{shortarticle}{true}
\usepackage{graphicx}
\usepackage{dcolumn}
\usepackage{bm}
\usepackage[]{siunitx}
\usepackage[utf8]{inputenc}
\usepackage[T1]{fontenc}
\usepackage{mathptmx}
\usepackage{etoolbox}
\usepackage{color}
\usepackage{lineno}
\usepackage[normalem]{ulem}
\usepackage{tabularx}
\usepackage{comment}
\title{Multidetection scheme for Transient-Grating-based spectroscopy}

\author[1,2]{M. Brioschi}
\author[1,2]{P. Carrara}
\author[2]{V. Polewczyk}
\author[2,3]{D. Dagur}
\author[2]{G. Vinai}
\author[2]{P. Parisse}
\author[2]{S. Dal Zilio}
\author[2]{G. Panaccione}
\author[1]{G. Rossi}
\author[2,*]{R. Cucini}

\affil[1]{Dipartimento di Fisica, Università di Milano, via Celoria 16, 20133 Milano, Italy}
\affil[2]{CNR-IOM, Strada Statale 14, km 163.5, 34149 Basovizza (TS), Italy}
\affil[3]{Dipartimento di Fisica, Università degli Studi di Trieste, Via Valerio 2, 34127 Trieste, Italy}
\affil[*]{Corresponding author: cucini@iom.cnr.it}

\begin{abstract}
Time-resolved optical spectroscopy represents an effective non-invasive approach to investigate the interplay of different degrees of freedom, which plays a key role in the development of novel functional materials. Here, we present magneto-acoustic data on Ni thin films on SiO$_2$ as obtained by  a versatile pump-probe setup that combines transient grating spectroscopy with time-resolved magnetic polarimetry. The possibility to easily switch from pulsed to continuous-wave probe allows probing of acoustic and magnetization dynamics on a broad timescale, in both transmission and reflection geometry.
\end{abstract}

\setboolean{displaycopyright}{true}

\begin{document}
\maketitle

\section{Introduction} \label{intro}
The challenge of material science and engineering imposes characterizing the low energy properties of samples to identify emerging technology opportunities. This requires several experimental techniques, as well as managing the same sample environment and experimental conditions when measuring different complementary properties in all-optical setups, and transferring samples to other experimental stations. Maximum integration of techniques in a given setup provides unique opportunities to acquire data that enable consistent analysis, reducing the uncertainties about the actual sample conditions. This approach is already implemented in large-scale facilities, e.g. in the NFFA suite connected with synchrotron radiation beamlines at Elettra \cite{apeHE}, but it is not yet common in table-top experiments. Transient Grating (TG) spectroscopy is a pump-probe technique allowing the investigation of systems driven out of equilibrium by optical triggers. Dynamical properties can be investigated by acquiring the time-resolved (tr) relaxation towards the ground state, with suitable detection schemes depending on the specific addressed phenomena (e.g. tr-polarimetry in magnetoacoustics studies \cite{janu2016,carrara2022}). Analysis of such tr-signals also provides insight into equilibrium properties, such as thermal diffusion coefficients, acoustic phonon spectrum, and magnetostatic parameters \cite{rogers2000, Cucini2011,carrara2022}. Here, we report on a versatile TG experimental setup developed at NFFA-SPRINT laboratory \cite{SPRINT2022} and available also as a user facility. Using complementary probes, the setup enables the investigation of acoustic and thermal phenomena on a broad time interval (from hundreds of fs to ms). Furthermore, the implementation of tr-polarimetry both in transmission and reflection makes this setup suitable for magnetoacoustic studies, on both transparent and opaque materials. We have tested the performance of this versatile setup by characterizing a 40-nm polycrystalline Ni film that is a prototypical system for magnetoacoustics. (see e.g. \cite{carrara2022, dreher2012surface}).
\section{Methodology}
The dynamical properties of a system can be efficiently probed using pump-probe spectroscopy, with a pump pulse exciting selected low-energy modes of the sample, and a delayed probe pulse measuring the relaxation of the excited degrees of freedom towards the ground state, giving access to the nonlinear properties of the material. Four Wave Mixing (FWM) is indeed a non-linear effect in which three pulses (two pumps and one probe), with the same or different wavelengths, interact with the medium and drive a third-order nonlinearity in the polarization acting as the source of a fourth beam, which contains information on the excited modes \cite{mukamel}. Several techniques are based on the FWM effect, each one exploiting different relationships between the pulse wavevectors (the so-called phase matching condition). In the TG case, two coherent pump pulses overlap on the sample surface, generating an interference grating. The line spacing $\Lambda$ of the grating is defined by the off-normal impinging angle $\theta_\text{pump}$ and the pump pulse wavelength $\lambda_\text{pump}$, and it is related to the induced wavevector \textbf{q} via $\lvert \textbf{q} \rvert=2 \pi/ \Lambda = 4 \pi \sin{\left( \theta_\text{pump}/2\right)}/\lambda_\text{pump}$ \cite{eichler1977laser}. The spatially-modulated pump light triggers a \textbf{q}-selective excitation-deexcitation chain, whose details depend on the photon energy and on the microscopic properties of the material \cite{eichler1977laser,choudhry2021characterizing}. 
In the case of parallel $s$-polarization for the two pumps, the induced intensity grating launches, in absorbing materials, two counter-propagating surface acoustic waves (SAWs) via thermoelastic expansion, whose superposition creates a transient standing-wave pattern.
The probe beam impinges on the sample and is diffracted by the grating. The diffracted intensity, at retarded times, provides information on the relaxing TG and so on the dynamical properties of the excited modes.
Various interaction channels can be excited, providing several information, like sound velocity and acoustic damping time in complex liquids \cite{cucini2010} as well as in thin films \cite{rogers2000}, structural and thermal properties \cite{Nelson2000}, spin dynamics \cite{carter2006} and magnetoelastic coupling \cite{janu2016, carrara2022}.

\begin{figure*}
\vspace{-0.2cm}
 \includegraphics[width=\textwidth]{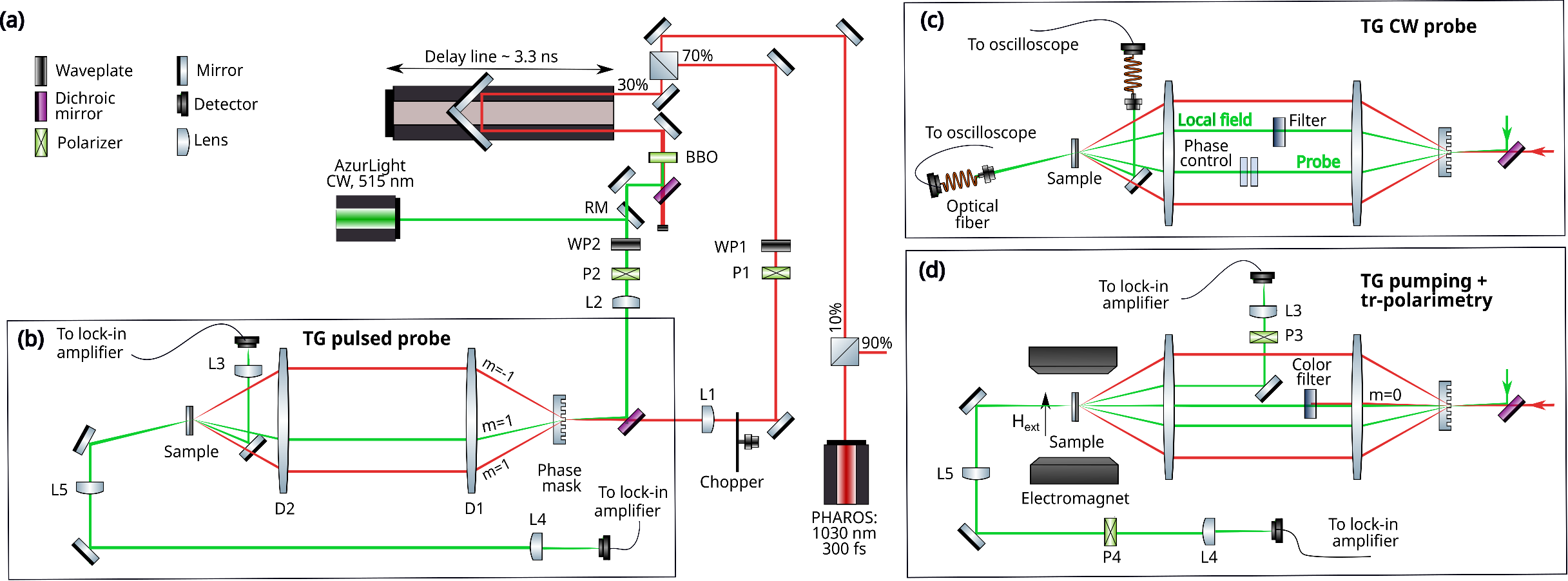}%
 \caption{(a) Laser sources and main body of the TG setup. (b) and (c) endstation of TG spectroscopy with pulsed and CW probe, respectively. (d) endstation for magnetoelastic measurements combining the TG pumping mechanism with tr-polarimetry.}
 \label{fig:setup}
 \end{figure*}
 \begin{figure}
\centering
\vspace{-0.7cm}
 \includegraphics[width=0.48\textwidth]{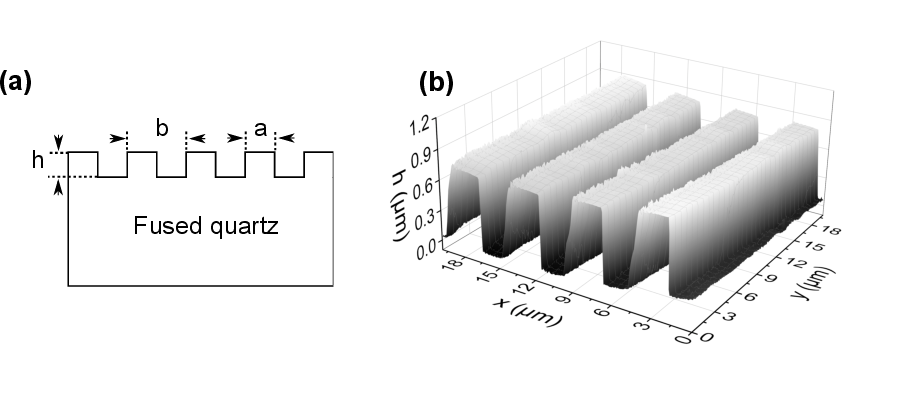}%
 \vspace{-0.5cm}
 \caption{(a) Phase mask sketch. (b) Phase mask characterization via AFM.}%
 \label{fig:pm}
 \end{figure}  
\section{The experimental setup} \label{tgsetup}
This section is organized as follows: the general aspects of the setup main body and of the TG pumping scheme are presented here below; then the optical setup at the sample endstation is presented for the pulsed probe (Section \ref{pulsedtgsetup}), for the continuous-wave (CW) probe (Section \ref{tgcw}) and for the tr-polarimetry (Section \ref{tgpol}). The possibility of keeping the same pumping mechanism and switching from pulsed to CW probe allows the investigation of the TG-induced dynamics from hundreds of fs to ms; each endstation setup is conceived to allow both reflection and transmission geometry. \newline
The pulsed laser source is a Yb:KGW-based integrated laser system (PHAROS, Light Conversion), characterized by turn-key operation and by high pulse-to-pulse stability. The laser provides infrared (IR) pump pulses at variable repetition rate in the tens-to-hundreds kHz range, with average power of 20 W, and energy per pulse up to $400$ \si{\micro \joule}. Considering that a fraction of \si{\micro \joule} is enough to establish the TG, we mostly work in parasitic mode so that the NFFA-SPRINT beamline can feed simultaneously other setups \cite{cucini2020coherent}. Typically, only $10 \%$ of the PHAROS output is sent to the TG setup. The IR beam is then split by a 70/30 beamsplitter: $70 \%$ is used for the pump while the remaining $30\%$ passes through a 2-mm thick Beta-Barium Borate (BBO) crystal, generating the second harmonic at $515$ \si{\nano \meter}, used as the pulsed probe. The maximum achievable time delay between pump and probe beams is approximately 3.3 ns, set through a corner-cube retroflector mounted on a 50-cm-long delay line. A dichroic mirror removes the residual part of the IR beam from the probe path. 
The switching from pulsed to CW probe only requires the insertion of a removable mirror (RM) routing the output of an additional single-mode fiber laser source (ALS, from AzurLight), which provides a 515-\si{\nano \meter} beam with tunable power up to 2 \si{\watt}. 
The chosen wavelength guarantees the possibility to use the same transmissive optical components (e.g. doublets, lenses) as for the second harmonic of the $1030$-\si{\nano \meter} PHAROS output. To finely tune the fluence on the sample, a half-wave plate followed by a Glan-laser polarizer is placed along both the pump (WP1, P1) and probe (WP2, P2) branches.
The pump and probe are collinearly coupled by a dichroic mirror and focused on a fused quartz phase mask by plano-convex lenses (L1, L2) with focal length $f = 20 $ \si{\centi \meter}. The phase masks are micromachined at the IOM-CNR laboratory \cite{lito} using combined UV lithography and dry etch processing. Following Reference \cite{meshalkin2019analysis}, the phase masks have the required efficiency on the $m=\pm1$ orders for 1030-\si{\nano \meter} and $515$-\si{\nano \meter} beams if $D = a/b \approx 0.5$ and $h \approx 750$ \si{\nano \meter}, where $D$ is the square-wave duty cycle and $h$ is the groove depth (see Figure \ref{fig:pm}.a). We fabricated three phase masks, with $b = 5.05$, $6.01$, and $8.00$ \si{\micro \meter}; Figure  \ref{fig:pm}.b shows the Atomic Force Microscopy (AFM) characterization of the phase mask with a $5.05$ \si{\micro \meter} periodicity. AFM scans are obtained in standard air conditions with a Solver Pro (NT-MDT) instrument, in semicontact mode using commercial cantilevers (NT-MDT, NSG30, nominal spring constant $k = 40 $ Nm$^{-1}$, nominal radius of curvature $ 10$ \si{\nano \meter}).\newline
The diffraction maxima generated by the phase mask are collimated and recombined by two custom achromatic and aspheric doublets (D1, D2) with focal length $f = 10$ cm, which guarantee the spatial overlap of both wavelengths at the sample position, also for beams impinging close to the doublet edges. The system is aligned in confocal configuration and the minimum spot size on sample is approximately $40$ \si{\micro \meter} in diameter. The TG is generated by interference of the $m = \pm 1$ diffraction maxima of the $1030$-\si{\nano \meter} beam.

\subsection{TG setup with pulsed probe} \label{pulsedtgsetup}
\begin{figure*}[t]
\vspace{-0.65cm}
 \centering \includegraphics[width=0.9\textwidth]{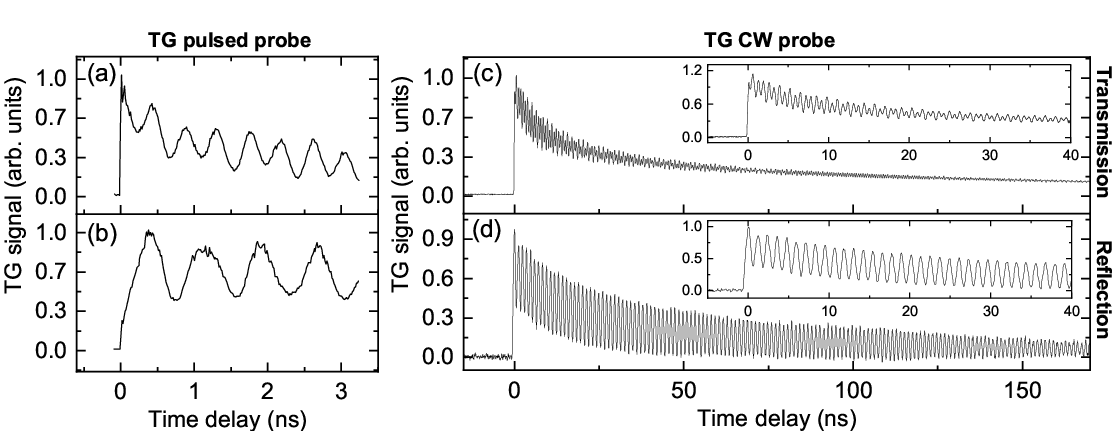}%
 \vspace{-0.1cm}
 \caption{TG acoustic signals for 40-nm Ni on SiO$_2$. TG signal with pulsed probe \textbf{(a)} in transmission and \textbf{(b)} in reflection geometry. TG signal with CW heterodyne probe \textbf{(c)} in transmission and \textbf{(d)} in reflection geometry. The signals are obtained via heterodyne amplification by subtracting traces with $\pi$ phase difference. The insets show the heterodyne signals in the first 40 \si{\nano \second}: the excitation of both acoustic modes is clearly observed in transmission.}
 \label{fig:tgacu}
 \end{figure*}

Figure \ref{fig:setup}.b shows the detection scheme for the TG setup with pulsed probe. In this configuration the fastest accessible dynamics is 300 \si{\femto \second}, limited by the pump pulse duration in the present instrument. The maximum time delay between pump and probe beams is approximately 3.3 \si{\nano \second}, as fixed by the delay line. The probe is the 515-nm first diffraction order ($m=+1$) of the second harmonic of the pump. Acoustic modes of the sample can be directly characterized by measuring the intensity of the probe, partially diffracted by the induced grating at the phase matching angle, in both transmission and reflection. The TG signal is focused by a plano-convex lens ($f=10$ cm) into a Femtowatt photoreceiver (2151 Newport), whose output is fed into a lock-in amplifier (SR860 from Stanford Research Systems); the data are acquired through a homemade LabVIEW software. In reflection geometry, the sample is tilted downwards, in order to separate the probe and the diffracted beam: due to the phase matching condition, the two beams have the same propagation axis (the diffracted beam in reflection geometry is drawn at a different angle in Fig. \ref{fig:setup}.a only for illustrative purposes). When working in transmission geometry, the two beams do not overlap and the sample surface plane can be set orthogonal to the scattering plane, thus ensuring more efficient spatial overlap and a smaller beam footprint.

\subsection{TG with continuous-wave probe} \label{tgcw}
For the CW setup (see Figure \ref{fig:setup}.c) we implemented heterodyne detection, which is commonly used in TG experiments \cite{cucini2010, rogers2000}: the diffracted probe is superimposed to a local field, which is used to amplify the signal. In our case, the superposition is guaranteed by using the $m=\pm 1$ diffraction orders of $515$-nm beam as probe and local field, respectively. The heterodyne detection requires a fine-tuning of the relative phase between the probe and the local field, which can be achieved through two custom-made prisms, designed to avoid any significant beam displacement. A filter is placed on the local field to avoid the saturation of the detector when higher probe fluence is needed. The signal is then focused into an optical fiber and it is measured by a high-speed detector (1544-B from Newport, 12 GHz bandwidth) connected to a digital oscilloscope (Lecroy, 4 GHz bandwidth). The acquisition chain limits the fastest accessible dynamics to a fraction of ns; however, it is suitable to investigate longer dynamics (up to ms), since the time interval is set by the oscilloscope.

\subsection{TG setup with time-resolved polarimetry} \label{tgpol}
Based on the pulsed-probe setup (see Section \ref{pulsedtgsetup}), we implemented tr-Faraday and tr-MOKE polarimetry (see Figure \ref{fig:setup}.d): this is a suitable detection solution  to study e.g. magnetoelastic coupling in the time domain. The time resolution and maximum observation window are 300 \si{\femto \second} and 3.3 \si{\nano \second}, respectively. We use the $m = 0$ and the $m = 1$ diffraction order of the 515-nm beam as probes for the tr-Faraday and tr-MOKE polarimetry, respectively. A color filter on the $m = 0$ removes the fundamental at 1030 nm, which might induce additional dynamics on the sample and/or unwanted heating. Upon interaction with the sample magnetization, the probe changes its polarization state, which is detected setting a polarizer (P3 and P4 in Figure \ref{fig:setup}.c for MOKE and Faraday effect, respectively) close to extinction. The signal acquisition chain (photoreceiver and lock-in amplifier) is the same as described in Section \ref{pulsedtgsetup}. Magnetic field is applied in the plane of the sample surface via a compact electromagnet, with maximum magnetic field of 100 \si{\milli \tesla}; the mounting of the electromagnet allows full in-plane azimuthal rotation and out-of-plane tilting up to 20° off-normal.
\section{Experimental results}
\begin{figure}
\vspace{-0.2cm}
\includegraphics[width=0.43\textwidth]{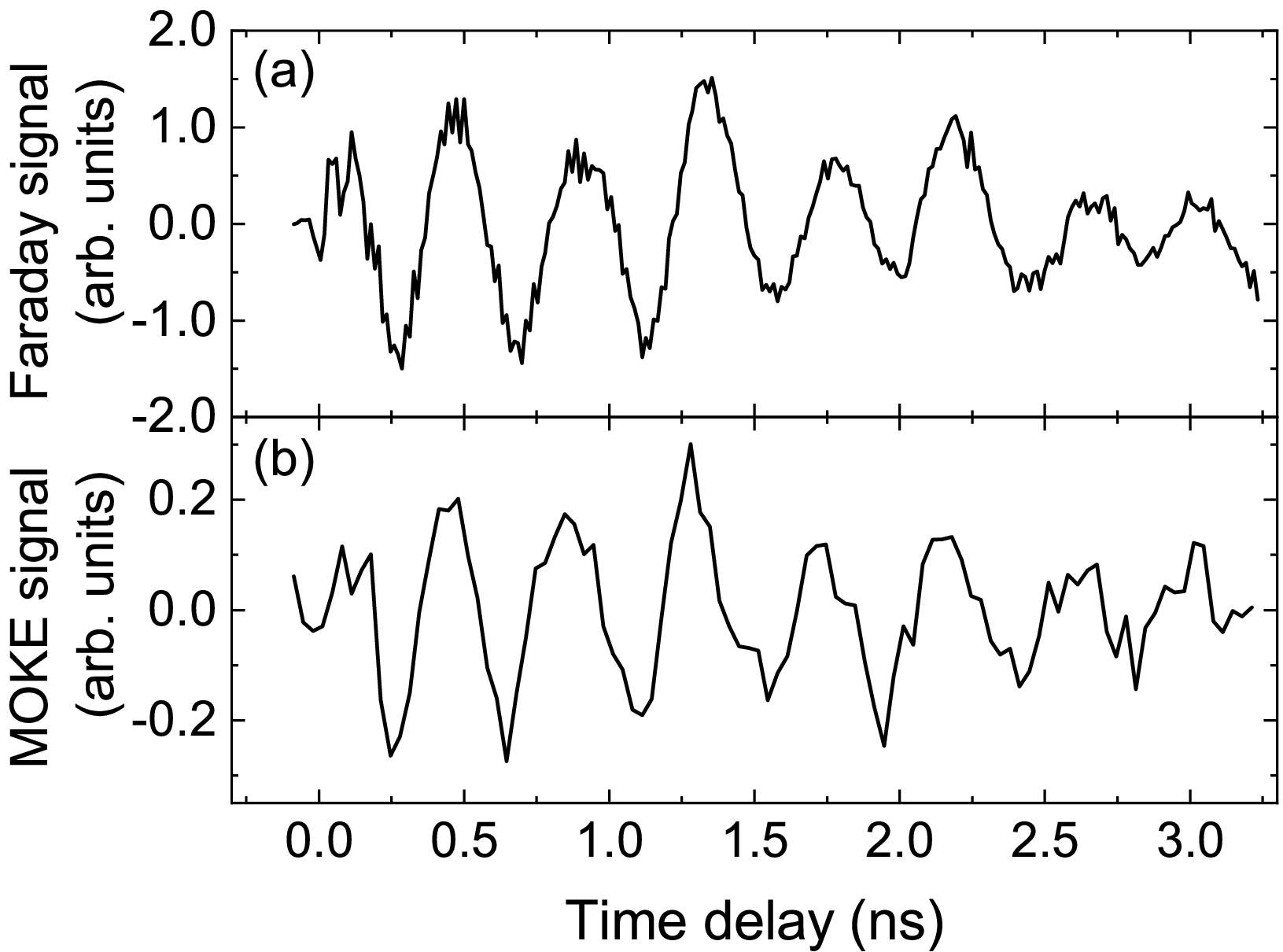}%
\vspace{-0.1cm}
 \caption{\textbf{(a)} TG-pumped tr-Faraday (transmission geometry) signal of a 40-nm Ni on SiO$_2$, under a magnetic field of 23 \si{\milli \tesla}. \textbf{(b)} Corresponding tr-MOKE signal (reflection geometry).}%
 \label{fig:MOKEfar}
 \end{figure}

Here, we present results on a 40-nm-thick Ni thin film grown on a SiO$_2$ substrate and capped with an 8-nm-thick SiO$_2$ film to prevent Ni oxidation. Details on growth and magnetic characterization are available in the Supplemental Material. The presented measurements, obtained with the three setups described in Section \ref{tgsetup}, are performed in air at room temperature.
Figure \ref{fig:tgacu} shows the acoustic data with pulsed or continuous probes,  both in transmission and reflection geometry. Consistently with previous work \cite{janu2016}, we detect two types of SAWs from transmission data: the Rayleigh SAW (RSAW) and the Surface Skimming Longitudinal Wave (SSLW). However, reflection data show only the RSAW. In the measurements obtained with the pulsed probe (Figure \ref{fig:tgacu}.a and b), we fixed the TG pitch $\Lambda=2.54$ \si{\micro \meter} ($q=2.47$ rad/\si{\micro \meter}) to maximize the number of oscillations, that are limited by the finite length of the delay line. In the case of CW probe (Figure \ref{fig:tgacu}.c and d), where the number of oscillations increases, we chose the smaller wavevector ($q=1.57$ rad/\si{\micro \meter}, corresponding to a TG pitch $\Lambda=4.00$ \si{\micro \meter}), in order to reduce the acoustic damping \cite{baggioli2022theory}. From the fast Fourier transform of the acoustic signals, we extract the frequency $\nu$ of the acoustic waves to calculate their phase velocity $c=2\pi \nu/q$.  
\begin{table}[t]
\centering
\vspace{-1cm}
\caption{\bf Sound velocity of SAWs extracted from the acoustic characterization on different timescales.}
\begin{tabular}{lcc}
\hline
 &  RSAW (\si{\kilo \meter}/\si{\second}) &SSLW (\si{\kilo \meter}/\si{\second}) \\
\hline
Pulsed, Trans.& $3.35 \pm 0.64$ & $5.96 \pm 0.64$  \\
Pulsed, Refl.& $3.34 \pm 0.61$ & --- \\
Continuous, Trans.& $3.39 \pm 0.05 $ & $6.05 \pm 0.05$\\
Continuous, Refl.& $3.37 \pm 0.03$ & --- \\
\hline
\vspace{-0.5cm}
\end{tabular}\label{tab:tablevelocity}
\end{table}
The sound velocities derived in the two probe schemes (see Table \ref{tab:tablevelocity}) are fully consistent. As expected, the sound velocities extracted from the data of the CW setup have a remarkably smaller error given the possibility of acquiring data on longer timescales. This is why CW transient grating setup is particularly suitable for thermal and acoustic studies \cite{rogers2000, cucini2010, dennett2018thermal}. In magnetic materials, the magnetization dynamics occur at much shorter times (fs to ns, \cite{ksenzov2021nanoscale}) and in-situ acoustic characterization on the very same timescale can  provide useful insights into magnetoelastic signals \cite{janu2016,carrara2022}. Figure \ref{fig:MOKEfar} shows the tr-Faraday and tr-longitudinal MOKE signals obtained by exploiting the TG-pumping mechanism and using an additional probe beam to detect the out-of-plane component of the magnetization. Unlike the  acoustic case, the signals in transmission and reflection geometry have the same frequency. This can be understood in the framework of magnetoelastic resonances \cite{carrara2022}. The transient grating technique allows to fix the wavevector $q$ of both the RSAW and SSLW; by tuning the external magnetic field so that the Kittel frequency matches either the RSAW or SSLW ones, acoustically-driven ferromagnetic resonance is reached and large out-of-plane component of the magnetization is observed during precession \cite{janu2016, carrara2022}. In Figure \ref{fig:MOKEfar}, Faraday and MOKE signals are shown, under a magnetic field of 23 \si{\milli \tesla}: at this field, the magnetization precession frequency matches the SSLW acoustic frequency. Static magnetic properties can be extracted via analysis of the resonance condition \cite{janu2016,carrara2022, vinci2017}. The signals are shown after background removal and in arbitrary units; the MOKE signal is about one order of magnitude smaller than the Faraday one, consistently with the off-normal MOKE incidence.
\section{Conclusions}
We described a versatile table-top setup that combines TG spectroscopy with tr-polarimetry, in transmission and reflection geometry. The main novelty is the possibility to perform TG measurements on complementary timescales, thanks to the use of both pulsed and CW laser  probes. This makes the setup optimal for both fine acoustic characterization and to study ultrafast phenomena in magnetoelastic studies, where it is key to probe simultaneously the acoustic and magnetization dynamics. Keeping the same TG-pumping mechanisms, we have reported tr-Faraday and tr-MOKE signals due to acoustically-driven ferromagnetic resonance. 
This is an important feature since previous reports are based only on transmission measurements. Here we have chosen a reference sample, which is transparent enough to observe tr-polarimetry signals in both reflection and transmission; however, for thicker and/or more opaque samples time-resolved MOKE might be the only way to access information about the magnetization dynamics. The various time scales available with our setup, and the possibility to access different degrees of freedom, enable to perform cutting-edge research in many fields, e.g. magneto-elastic coupling in magnetic thin films and magnonic crystals, wavevector-selective exciton dynamics in 2D materials \cite{wang2019diffusion} and (biocompatible) ferrofluids \cite{bacri1995transient}. Finally, the TG setup will be used also for testing samples and providing reference results for complementary dynamics studied with Free Electron Laser sources \cite{ksenzov2021nanoscale}, where higher wavevectors are accessible.  

\begin{backmatter}
\bmsection{Funding} This work has been performed in the framework of the Nanoscience Foundry and Fine Analysis (NFFA-MUR Italy Progetti Internazionali) facility, and of the NFFA-Europe-Pilot under H2020 grant agreement n. 101007417.
\bmsection{Acknowledgments} The authors thank A. Fondacaro for technical support.
\bmsection{Disclosures} The authors declare no conflicts of interest.
\bmsection{Data Availability Statement}{Data underlying the results presented in this paper are not publicly available at this time but may be obtained from the authors upon reasonable request.}
\bmsection{Supplemental document}
See Supplement 1 for supporting content. 
\end{backmatter}

\end{document}